\documentclass[11pt]{article}
\usepackage{graphicx,rotating,cite,epsfig,amsmath,amssymb}

\def\q2{$Q^2$}

\def\gev2{$~\mbox{GeV}^2$}

\font\tenrm=cmr10

 1
 1
 1
 3
 2
 2
 2

\begin{document}

\begingroup
\thispagestyle{empty}
\baselineskip=14pt
\parskip 0pt plus 5pt
\begin{center}
\end{center}

\begin{flushright}
OPERA Internal Note \\
\today
\end{flushright}
\bigskip

\begin{center}
  \huge \textbf{Sensitivity to $\theta_{13}$ of the CERN to Gran Sasso neutrino
    beam}
\end{center}

\bigskip\bigskip
\begin{center}
{\Large M.~Komatsu$^1$, P.~Migliozzi$^2$ and F.~Terranova$^3$} \\
$^1$~Nagoya University, Japan \\
$^2$~INFN, Napoli, Italy \\
$^3$~INFN, Laboratori Nazionali Frascati, Italy \\
\end{center}

\begin{abstract}
The sensitivity of the present CNGS beam to the sub-dominant $\nu_\mu
\leftrightarrow \nu_e$ oscillations in the region indicated by the
atmospheric neutrino experiments is investigated. In particular, we
present a revised analysis of the OPERA detector and discuss the
sensitivity to $\theta_{13}$ of ICARUS and OPERA combined. We show
that the CNGS beam optimised for $\nu_\tau$ appearance, will improve
significantly (about a factor 5) the current limit of CHOOZ and
explore most of the region
$\sin^22\theta_{13}\simeq\mathcal{O}(10^{-2})$.
\end{abstract}

\vskip 0.4cm
\vfill
\begin{center}
\end{center}

\endgroup

\newpage
\pagestyle{plain}
\pagenumbering{arabic}
\setcounter{page}{2}
\newpage
\baselineskip=14pt 
\tenrm

%
%

\section{Introduction}
\label{introduction}

There are currently intense efforts to confirm and extend the evidence
for neutrino oscillations in the atmospheric and solar sectors. In
particular the disappearance of muon neutrinos originating from cosmic
ray interactions~\cite{Fukuda:1994mc,Fukuda:1998mi} provides a strong
hint in favour of $\nu_\mu \leftrightarrow \nu_\tau$
oscillations. This indication will be tested in a conclusive manner at
the next generation of long baseline experiments.  In a simple
two-flavour mixing scheme the atmospheric neutrino experiments
constrain the mixing parameter in the region around ($\Delta m^2$,\,
$\sin^2 2\theta$)=($2.5 \times 10^{-3}$,\, 1). Recently, the SNO
experiment gave an unambiguous signature of the solar neutrino
oscillations, excluding the sterile neutrino
hypothesis~\cite{Ahmad:2002ka,Ahmad:2002jz}.

In addition to the dominant $\nu_\mu \leftrightarrow \nu_\tau$
component, sub-dominant oscillations involving also $\nu_e$ could be
observed.  In this three-flavour scenario the gauge eigenstates
$\nu_\alpha$ ($\alpha=e,\mu,\tau$) are related to the mass eigenstates
$\nu'_i$ ($i=1,2,3$) through the leptonic mixing matrix $U_{\alpha
i}$.  This matrix can be parametrized as:

 \begin{equation}
U(\theta_{12},\theta_{13},\theta_{23},\delta)=\left(
\begin{tabular}{ccc}
$c_{12}c_{13}$      & $s_{12}c_{13}$   &  $s_{13}e^{-i\delta}$ \\
$-s_{12}c_{23}-c_{12}s_{13}s_{23}e^{i\delta}$ &
$c_{12}c_{23}-s_{12}s_{13}s_{23}e^{i\delta}$ & $c_{13}s_{23}$ \\
$s_{12}s_{23}-c_{12}s_{13}c_{23}e^{i\delta}$ &
$-c_{12}s_{23}-s_{12}s_{13}c_{23}e^{i\delta}$ & $c_{13}c_{23}$
\end{tabular}\right)                                                           \end{equation}
with $s_{ij}=\sin\theta_{ij}$ and $c_{ij}=\cos\theta_{ij}$. 

Here the leading $\nu_\mu \leftrightarrow \nu_\tau$ transitions are driven by
the angle $\sin^2 2\theta_{23}$ and the squared mass difference $\Delta
m^2_{23}$.  Sub-dominant $\nu_\mu \leftrightarrow \nu_e$ oscillations at the
atmospheric scale ($\Delta m^2_{23} \simeq 2.5 \times 10^{-3}$) are driven by
the mixing angle $\theta_{13}$. This angle is constrained by the CHOOZ
experiment to be small ($\sin^2 2\theta_{13} < {\cal O}
(10^{-1})$)~\cite{Apollonio:1999ae}.
 
The next generation of long baseline experiments --- MINOS~\cite{minos},
ICARUS~\cite{icarus} and OPERA~\cite{operaproposal} --- will search for a
conclusive evidence of the oscillation mechanism at the
atmospheric scale. In particular, the CNGS beam~\cite{CNGS,ngsaddendum}
will directly confirm the indication for neutrino oscillations through tau
appearance. 

Although the precision that will be achieved on the parameters driving the
dominant $\nu_\mu\leftrightarrow\nu_\tau$ after this first generation of long
baseline experiments is well established~\cite{barger}, the sensitivity to the
sub-dominant $\nu_\mu \leftrightarrow \nu_e$ channel deserves further
investigation, in particular for what concerns CNGS.  This is due to the fact
that only recently the baseline configuration of the CNGS detectors
(ICARUS~\cite{icarus} and OPERA~\cite{planopera}) has been established.
Moreover, a full $\nu_\mu \leftrightarrow \nu_e$ analysis with the OPERA
experiment has not been carried out yet.

In this paper we discuss the sensitivity of the ICARUS and OPERA experiments in
searching for $\theta_{13}$ by using the present CNGS neutrino beam. The
sensitivity of ICARUS to $\theta_{13}$ has been evaluated in~\cite{icarus}. For
OPERA, whose primary aim is the direct detection of $\nu_\tau$ appearance, a
preliminary estimate of the sensitivity to $\nu_\mu \leftrightarrow \nu_e$
oscillations has been given in the 1999 Letter of Intent~\cite{progrep} and is
evaluated here in more details. The paper is organised as follows: in
Section~\ref{cngs-program} we briefly describe the present CNGS neutrino beam,
the ICARUS and OPERA experimental setups and their features; in
Section~\ref{sensi} we discuss the ICARUS and OPERA sensitivity to
$\theta_{13}$
; finally in Section~\ref{conclu} we draw our conclusions.

\section{The present CNGS experimental program}
\label{cngs-program}

\subsection{The CNGS neutrino beam}
In November 2000 a new version of the CNGS beam design was
released~\cite{cngsbeam}. In Tables~\ref{tab:results1}
and~\ref{tab:results2}, the performance of the beam are given. \\
 
\begin{table}[hbtp]
  \small
\begin{center}
\caption{\small Nominal performance of the CNGS reference
beam~\cite{ngsaddendum}.}
\vspace{5mm}
\begin{tabular}{||c|c||}
\hline
\hline
&  \\
$\nu_{\mu}$\,(m$^{-2}$/pot)  &  7.78\,$\times$\,10$^{-9}$ \\
\hline
&  \\
$\nu_{\mu}$ CC events/pot/kton  & 5.85\,$\times$\,10$^{-17}$\\
\hline
&  \\
$\nu_e$/$\nu_{\mu}$         & 0.8\,\% \\
\hline
&  \\
$\overline{\nu}_{\mu}$/$\nu_{\mu}$  & 2.1\,\% \\
\hline
&  \\
$\overline{\nu}_e$/$\nu_{\mu}$    &0.07\,\%\\
\hline
\hline
\end{tabular}
\label{tab:results1}
\end{center}
\end{table}
 
\begin{table}[hbtp]
  \small
\begin{center}
\caption{\small Number of $\nu_\tau$ charged-current interactions at Gran Sasso per
  $kton$ and per year (shared mode) for different values of $\Delta m^2$ and
  $\sin^2(2\theta)\,=\,1$ \cite{ngsaddendum}.}  \vspace{5mm}
\begin{tabular}{||c|c||}
\hline
\hline
& \\
$\Delta m^2$ & $\nu_\tau$ CC interactions/kton/year \\
\hline                                                                         & \\
 1\,$\times$\,10$^{-3}$\,eV$^2$  &  2.53 \\
\hline
& \\
 3.\,$\times$\,10$^{-3}$\,eV$^2$ & 22.8 \\
\hline
& \\
 5\,$\times$\,10$^{-3}$\,eV$^2$ & 63.3 \\
\hline
\hline
\end{tabular}
\label{tab:results2}
\end{center}
\end{table}

The possibility of an increase of the neutrino beam intensity at
moderate cost is being studied by the CNGS design
group~\cite{cngsbeam} and other groups working on the PS and SPS
machines~\cite{ref4plan}. The SPS and its injectors have a crucial
role for the LHC accelerator and considerable efforts and investment
are going into the upgrade of these accelerators to meet the
requirements of LHC. Most of the efforts are also directly beneficial
to CNGS: for example, it is expected that the maximum intensity
accelerated in the SPS will be $7\times 10^{13}$ rather than
$4.5\times 10^{13}$ pot/cycle~\cite{ref5plan}. Moreover, it has been
recently pointed out~\cite{ref4plan} that, taking into account that
all the accelerators (LINAC, PSB, PS and SPS) are close to their
limits, an increase in the proton beam intensity can be reached -
pending further accelerator machine development studies - by reducing
the beam losses between two sequential accelerators. These
improvements could, at a moderate cost, increase the proton flux by as
much as a factor of ~1.5.

Other scenarios for a further increase of the SPS extracted beam
intensity has been outlined in~\cite{ref4plan}, but those are much
more expensive and might turn out to be technically unrealistic. It is
therefore prudent to state that an intensity increase for CNGS of a
factor 1.5 is within reach.

The CNGS energy spectrum has been optimized for $\nu_\tau$ appearance
and the prompt $\nu_\tau$ contamination (mainly from $D_s$ decays) is
negligible. On the other hand, in the CNGS beam, the expected $\nu_e$
contamination is relatively small compared to the dominant $\nu_\mu$
component, see Table~\ref{tab:results1} and allow to search for the
sub-dominant oscillation $\nu_\mu \leftrightarrow \nu_e$ seeking after
an excess of $\nu_e$ charged-current (CC) events.

The systematic error associated with the $\nu_e$ contamination plays
an important role for the $\nu_\mu\leftrightarrow\nu_{e}$ oscillation
search, the statistical fluctuation of this $\nu_e$ component being
the irreducible limiting factor. In turns, this uncertainty depends on
the knowledge of the $K$ yield. For sake of consistency with
Refs.~\cite{barger,icabeam}, we assume for this study a 5\% systematic
error on the overall $\nu_e$ flux.  However, as we will show in
Section~\ref{eve_sel} due to the small number of selected events,
especially in OPERA, a more conservative estimate of the systematics
(up to 10\%) would not change appreciably the experimental
sensitivity.

\subsection{The ICARUS experiment}

The aim of the ICARUS Collaboration~\cite{icarus1} is the construction of a
liquid argon time projection chamber (TPC) detector with a space resolution
comparable to that of bubble chambers.
 
In an ICARUS module a large liquid argon volume is contained in a
cryostat equipped with wire readout planes and a high voltage plane at
the center giving a long drift length. Ionisation electrons produced
by charged particles crossing the liquid argon are collected at the
readout planes. The arrival time of the different electrons allows the
reconstruction of the particle track trajectory by the time projection
technique.  Each readout plane has three coordinates at $60^\circ$
from each other and wire pitch of 3~mm. This corresponds to a
\textit{bubble} diameter of about 3~mm.

The detector is a homogeneous tracking device capable of $dE/dx$ measurement.
Its high $dE/dx$ resolution allows good momentum measurement and particle
identification of soft particles. Electromagnetic and hadronic showers are, in
fact, identified and fully sampled. This gives the energy resolution for
electromagnetic showers of $\sigma(E)/E=3\%/\sqrt{E(GeV)}\oplus1\%$ and for
hadronic contained showers of $\sigma(E)/E = 16\%/\sqrt{E(GeV)}\oplus1\%$.
ICARUS features excellent electron identification and $e/\pi^0$ separation.

The physics programme of the ICARUS experiment as described in the
Proposal~\cite{icarus1} has been derived assuming a ``baseline
configuration'' consisting of one T600 plus two T1200 modules for a
total fiducial mass of 2.35~kton~\cite{icarus}.

\subsection{The OPERA experiment}
\label{opera}
 
The OPERA experiment~\cite{operaproposal,operastatus} is meant for a long
baseline search for $\nu_{\mu}\leftrightarrow\nu_{\tau}$ oscillations in the
CNGS beam~\cite{cngsbeam}.  The experiment uses nuclear emulsions as high
resolution tracking devices for the direct detection of the $\tau$ produced in
the $\nu_{\tau}$ CC interactions with the target. Due to its excellent electron
identification capability exploited for the reconstruction of the $\tau
\rightarrow e$ decay, OPERA is able to perform also a $\nu_\mu \leftrightarrow
\nu_e$ oscillation search.

OPERA is designed starting from the Emulsion Cloud Chamber (ECC)
concept (see references quoted in \cite{operaproposal}) which combines
the high precision tracking capabilities of nuclear emulsions and the
large mass achievable by employing metal plates as a target. The basic
element of the ECC is the cell which is made of a 1~mm thick lead
plate followed by a thin emulsion film.  The film is made up of a pair
of emulsion layers $50~\mu$m thick on either side of a $200~\mu$m
plastic base.  Charged particles give two track segments in each
emulsion film.  The number of grain hits in $50~\mu$m ($15$-$20$)
ensures redundancy in the measurement of particle trajectories.

For details on the event reconstruction both with the nuclear emulsions and the
electronic detector, and the OPERA sensitivity to
$\nu_{\mu}\leftrightarrow\nu_{\tau}$ we refer
to~\cite{operaproposal,operastatus}. The mass of the OPERA ``baseline
configuration''is 1.77~kton~\cite{planopera}. In the following an effective
mass of 1.65~kton has been assumed to account for the brick removal during data
taking.

Thanks to the dense ECC structure and the high granularity provided by the
nuclear emulsions, the OPERA detector is also suited for electron and $\gamma$
detection~\cite{operaproposal}.  The resolution in measuring the energy of an
electromagnetic shower in the energy range relevant to CNGS is approximately
constant and is about 20\%. Furthermore, the nuclear emulsions are able to
measure the number of grains associated to each track. This allows an excellent
two tracks separation (better than $1~\mu$m).  Therefore, it is
possible to disentangle single-electron tracks from tracks produced by electron
pairs coming from $\gamma$ conversion in the lead. These features are
particularly important for the $\nu_\mu\leftrightarrow\nu_e$ analysis.

The outstanding position resolution of nuclear emulsions can also be
used to measure the angle of each charged track with an accuracy of
about 1~mrad. This allows momentum measurement by using the Multiple
Coulomb Scattering with a resolution of about 20\% and the
reconstruction of kinematical variables characterising the event
(i.e. the missing transverse momentum at the interaction vertex
$p_T^{miss}$ and the transverse momentum of a track with respect to
hadronic shower direction $Q_T$).

\section{Sensitivity to $\theta_{13}$}
\label{sensi}

\subsection{Event selection}
\label{eve_sel}

In this Section we discuss the event selection in OPERA and the expected number
of signal and background events in both ICARUS and OPERA.

The overall rate is defined as

\begin{equation}
R(\Delta m^2_{23},\sin^2 2\theta_{23},\sin^2 2\theta_{13}) =
S+B^{(\tau\rightarrow e)}+ B^{(\nu_\mu \mbox{CC}\rightarrow\nu_\mu \mbox{NC})}+
B^{(\text{beam})}+B^{\text{(NC)}}\,,
\label{mu-el-minos}
\end{equation}
where the signal, $S$, is the convolution of the $\nu_\mu$ flux
($\phi_{\nu_{\mu}}(E)$) with the $\nu_e$ charged-current cross section
($\sigma_{\nu_{e}}^{\text{CC}}(E)$), the $\nu_\mu\leftrightarrow\nu_e$
oscillation probability ($P_{\nu_{\mu} \leftrightarrow \nu_{e}}(E)$)
and the the signal efficiency ($ \epsilon^{\mbox{signal}}$)

\begin{eqnarray}
S = A \int   \phi_{\nu_{\mu}}(E) \, P_{\nu_{\mu} \leftrightarrow
\nu_{e}}(E) \sigma_{\nu_{e}}^{\text{CC}}(E)\, \epsilon^{\mbox{signal}}  \, dE\,.
\end{eqnarray}

 The normalisation $A$ takes into account the effective target
mass. The background coming from the dominant $\nu_\mu \rightarrow
\nu_\tau$ oscillation channel where the produced $\tau$ decays in one
electron is
\begin{eqnarray}
B^{({\tau\rightarrow e})} = A \int   \phi_{\nu_{\mu}}(E) \,
P_{\nu_{\mu}\leftrightarrow \nu_{\tau}}(E)\sigma_{\nu_{\tau}}^{\text{CC}}(E) \,
\epsilon^{({\tau\rightarrow e})} 
dE \,,
\end{eqnarray}

where $\epsilon^{({\tau\rightarrow e})}$ is the efficiency and
$\sigma_{\nu_{\tau}}^{\text{CC}}$ is the $\nu_\tau$ charged-current
cross-section. This background includes mainly taus
decayed in the same lead plate as the primary vertex (``short
decays'') plus a small (4\%) contribution from ``long decays'' where
the kink has not been detected.

 The background from $\nu_\mu$ events with the primary muon not
identified is given by

\begin{eqnarray}
B^{(\nu_\mu \mbox{CC}\rightarrow \nu_\mu \mbox{NC})} = A \int   \phi_{\nu_{\mu}}(E) \, P_{\nu_{\mu} \leftrightarrow
\nu_{\mu}}(E) \sigma_{\nu_{\mu}}^{\text{CC}}(E)
\epsilon^{(\nu_\mu\mbox{CC}\rightarrow\nu_\mu\mbox{NC})} dE &,
\end{eqnarray}

where $\epsilon^{({\nu_\mu\mbox{CC}\rightarrow\nu_\mu\mbox{NC}})}$ is
the probability for a $\nu_\mu$CC to be identified as a $\nu_\mu$NC
and with another track mimicking an electron. 

The background coming from the $\nu_e$ beam contamination is
\begin{eqnarray}
B^{(\text{beam})} =
A \int   \phi_{\nu_{e}}(E) \, P_{\nu_{e} \leftrightarrow \nu_{e}}(E) \sigma_{\nu_{e}}^{\text{CC}}(E) \, 
\epsilon^{(\text{beam})}dE &\, ,
\end{eqnarray}

where $\epsilon^{(\text{beam})}$ is the efficiency for CC events from the
$\nu_e$ beam contamination. 

The last background included, $B^{(\text{NC})}$, from the decay of
neutral pions created by neutral-current (NC) interactions, is
\begin{eqnarray}
B^{(\text{NC})} = A \int   \phi_{\nu_{\mu}}(E) \, \sigma_{\nu}^{\text{NC}}(E) 
\, \epsilon^{(\text{NC})}
dE &,
\end{eqnarray}
where $\epsilon^{(\text{NC})}$ is the reduction efficiency. Further background
from charge exchange process is negligible \cite{operaproposal}.

The OPERA analysis for $\nu_\mu\leftrightarrow\nu_e$ oscillation quoted in the
1999 Letter of Intent~\cite{progrep} has been revised. The major improvements
are the following:

\begin{itemize}
\item the muon identification is improved from 80\% to 95\%
  \cite{operaproposal}. This means that the contribution to the background from
  $\nu_\mu$CC events can drastically be reduced;
\item $\gamma$'s produced at the primary vertex ($\pi^0$ decays) and
converting into the lead plate where the interaction occurred are
identified if at least one of the below conditions is fulfilled
\begin{enumerate}
\item the $e^+e^-$ pair has an opening angle larger than 3~mrad. This
cut is fixed by the angular resolution achievable by the emulsion
films;
\item if the $e^+e^-$ pair has an opening angle smaller than 3~mrad,
we exploit the nuclear emulsion capability in measuring the number of
grains associated with a reconstructed track. The rejection power of
this cut has been computed by assuming for a minimum ionising particle
30~grains per emulsion sheet, while 60~grains are assumed for the
$e^+e^-$ pair with opening angle smaller than 3~mrad.  A reconstructed
track is classified as an $e^+e^-$ pair if the associated number of
grains is larger than ($30+3\times\sqrt{30}$). Conservatively we
consider only the emulsion film downstream from the vertex lead
plate. Note, however, that a significant improvement could be obtained
including the grains in the subsequent emulsion sheets, crossed by the
electrons before showering;
\end{enumerate}
\item since neutrino oscillations create an excess of $\nu_e$CC
induced events at low neutrino energies, a cut on the visible energy
is applied.
\end{itemize}

Hence, the OPERA $\nu_\mu \leftrightarrow \nu_e$ dedicated analysis
seeks for neutrino interactions with a candidate electron from the
primary vertex with an energy greater than 1 GeV (this cut reduces the
soft $\gamma$ component) and a visible energy of the event smaller
than 20~GeV (this cut reduces the prompt $\nu_e$ component of the
background). Moreover, we also apply a cut on the number of grains
associated with the track of the candidate electron. The latter has a
strong impact on the reduction of the background from $\nu_\mu$CC and
$\nu_\mu$NC events and allowed for a softer cut on the electron
energy.  Finally, a cut on missing transverse momentum at the primary
vertex $p_T^{miss}$ of the event is applied ($p_T^{miss}<1.5$~GeV) to further
reduce NC contaminations and suppress $\tau\rightarrow e$ background.

For the brick finding, the vertex finding, the trigger efficiencies
and the fiducial volume cut, we used the numbers quoted in the Status
Report~\cite{operastatus}. The product of these efficiencies is shown
in Table~\ref{tab:effopera} and is indicated with $\xi$.

The final efficiencies ($\epsilon$) for both the signal and the
background channels are shown in the second raw of
Table~\ref{tab:effopera} and the expected number of events is shown in
Table~\ref{tab:eventi}.

The search for $\nu_\mu\leftrightarrow\nu_e$ oscillations does not
increase the emulsion scanning load by a large amount. Indeed, once
the events have been located, the time needed to search for
interesting topologies around the reconstructed vertex is small
compared with the time needed to locate the events. On the other hand,
for events identified as NC produced in the downstream half of the
brick one should always remove the following brick to perform the
electron identification. It implies that 13\% more bricks have to be
removed and scanned.  Notice that in the preliminary estimate quoted
in the 1999 Letter of Intent only half of the ECC mass has been used.
 
\begin{table}[hbtp]
  \small
\begin{center}
\caption{\small Signal and background efficiencies for the
$\nu_\mu\leftrightarrow\nu_e$ oscillation search with OPERA. Notice
that the $\tau\rightarrow e$ efficiency includes the
$\tau\rightarrow e\nu\bar{\nu}$ branching ratio.}
\vspace{5mm}
\begin{tabular}{||c|c|c|c|c|c||}
\hline
\hline
 & $\nu_e$CC signal & $\tau\rightarrow e$ & $\nu_\mu\mbox{CC}\rightarrow\nu_\mu\mbox{NC}$ & $\nu_\mu$NC & $\nu_e$CC beam  \\
\hline
$\xi$ & 0.53 & 0.53 & 0.52 & 0.48 & 0.53 \\
\hline
$\epsilon$ & 0.31 & 0.032 & $0.34\times10^{-4}$ & $7.0\times10^{-4}$ & 0.082 \\
\hline
\hline
\end{tabular}
\label{tab:effopera}
\end{center}
\end{table}

If we assume that both $\nu_\mu\leftrightarrow\nu_\tau$ and
$\nu_\mu\leftrightarrow\nu_e$ oscillations occur simultaneously with
oscillation parameters $\Delta m_{23}^2
=2.5\times10^{-3}~\mbox{eV}^2$, $\theta_{23}=45^\circ$ and
$\theta_{13}\in [3^\circ\div 9^\circ ]$, the expected number of signal
and background events both in ICARUS and OPERA are given in
Table~\ref{tab:eventi}. The rates are normalised to 5 years data
taking and assuming the nominal intensity of the CNGS beam.
 
\begin{table}[hbtp]
  \small
\begin{center}
\caption{\small Expected number of signal and background events for
both ICARUS and OPERA assuming 5 years data taking  with the nominal CNGS beam and oscillation
parameters $\Delta m_{23}^2 =2.5\times10^{-3}~\mbox{eV}^2$,
$\theta_{23}=45^\circ$ and $\theta_{13}\in [3^\circ\div 9^\circ ]$.}
\vspace{5mm}
\begin{tabular}{||c|c|c|c|c|c|c||}
\hline
\hline
$\theta_{13}$ & $\sin^22\theta_{13}$ & $\nu_e$CC signal & $\tau\rightarrow e$ & $\nu_\mu\mbox{CC}\rightarrow\nu_\mu\mbox{NC}$ & $\nu_\mu$NC & $\nu_e$CC beam  \\
\hline
\hline
{\bf ICARUS} & & & & & & \\
\hline
\hline
$9^\circ$ & 0.095&27 & 24 & - & - & 50 \\
\hline 
$8^\circ$ & 0.076&21 & 24 & - & - & 50 \\
\hline 
$7^\circ$ & 0.058&16 & 24 & - & - & 50 \\ 
\hline 
$5^\circ$ & 0.030&8.4 & 25 & - & - & 50 \\ 
\hline 
$3^\circ$ & 0.011& 3.1 & 25 & - & - & 50 \\ 
\hline
\hline
{\bf OPERA} & & & & & &  \\
\hline
\hline
$9^\circ$ & 0.095 &9.3 & 4.5 & 1.0&5.2& 18 \\ 
\hline           
$8^\circ$ & 0.076 & 7.4& 4.5 & 1.0&5.2& 18 \\ 
\hline           
$7^\circ$ & 0.058 & 5.8 & 4.6 & 1.0&5.2& 18 \\ 
\hline           
$5^\circ$ & 0.030 & 3.0 & 4.6 & 1.0&5.2& 18 \\ 
\hline           
$3^\circ$ & 0.011 & 1.2 & 4.7 & 1.0&5.2& 18 \\ 
\hline 
\hline
\end{tabular}
\label{tab:eventi}
\end{center}
\end{table}

The ICARUS rates at CNGS from $\nu_\mu\leftrightarrow\nu_e$
oscillations and from background sources have been taken from
Table~5.9 of Ref.~\cite{icarus} and scaled at the different
oscillation parameters $(\Delta m^2_{32},\sin^22\theta_{23},\sin^2
2\theta_{13})\, .$ Here a cut on the electron energy ($E_e>1$~GeV) and
the visible energy ($E_{vis}<20$~GeV) have been applied. In
Table~\ref{tab:eventi} the background from $\pi^0$ production in
neutral-current events is negligible for ICARUS as assumed in
Ref.~\cite{icarus}. This is not true for OPERA being the lead
radiation length much shorter than the one of liquid
argon. Nevertheless, this background can be further suppressed by a
kinematical analysis, see Section~\ref{kine}. On the other hand, the
$\tau\rightarrow e$ background in OPERA is suppressed by the
capability of this detector to reconstruct exclusively the $\tau$
decay topology. Finally, notice that in both experiments the dominant
background remains the $\nu_e$ contamination of the beam.

\subsection{Kinematical analysis}
\label{kine}

\begin{figure}[htbp]
\centering
\epsfig{file=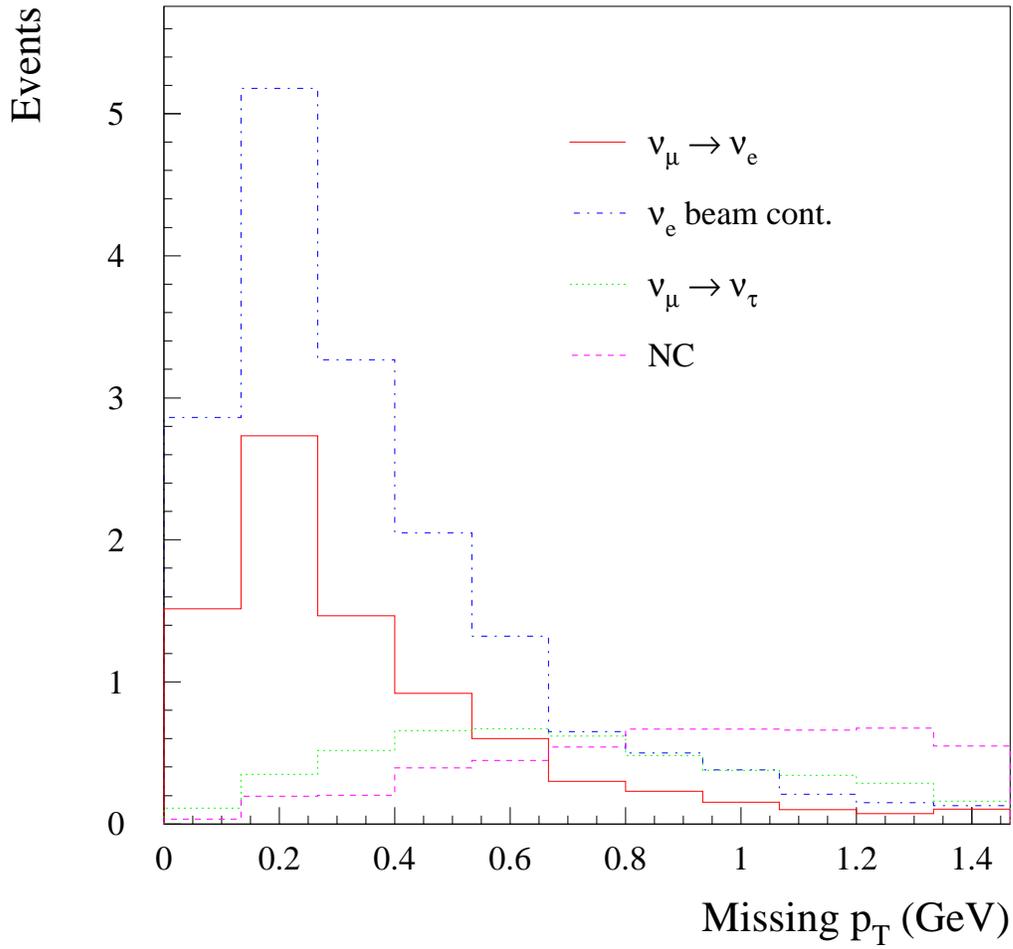,width=15cm}
\caption{ $p_T^{miss}$ distribution for signal and background channels
assuming $\theta_{13}=8^\circ$, $\Delta
m^2_{23}=2.5\times10^{-3}~\mbox{eV}^2$ and $\theta_{23}=45^\circ$.}
\label{fig:ptmiss}
\end{figure}

An increase of sensitivity to $\nu_\mu\leftrightarrow\nu_e$
oscillations can be obtained by fitting the kinematical distributions
of the selected events. As an example the OPERA $p_T^{miss}$
distribution for signal and background channels is given in
Fig.~\ref{fig:ptmiss}.

In order to disentangle $\nu_\mu\leftrightarrow\nu_e$ and
$\nu_\mu\leftrightarrow\nu_\tau$ oscillations, the ICARUS Collaboration
considered the transverse missing momentum. The $\nu_\mu\leftrightarrow\nu_e$
events are balanced, while $\nu_\mu\leftrightarrow\nu_\tau$ show a sizable
transverse momentum due to the presence of two neutrinos in the final state.
The visible energy and the missing transverse momentum are combined into a
binned $\chi^2$-fit, in which the three oscillation parameters are allowed to
vary.

\begin{figure}[htbp]
\centering
\epsfig{file=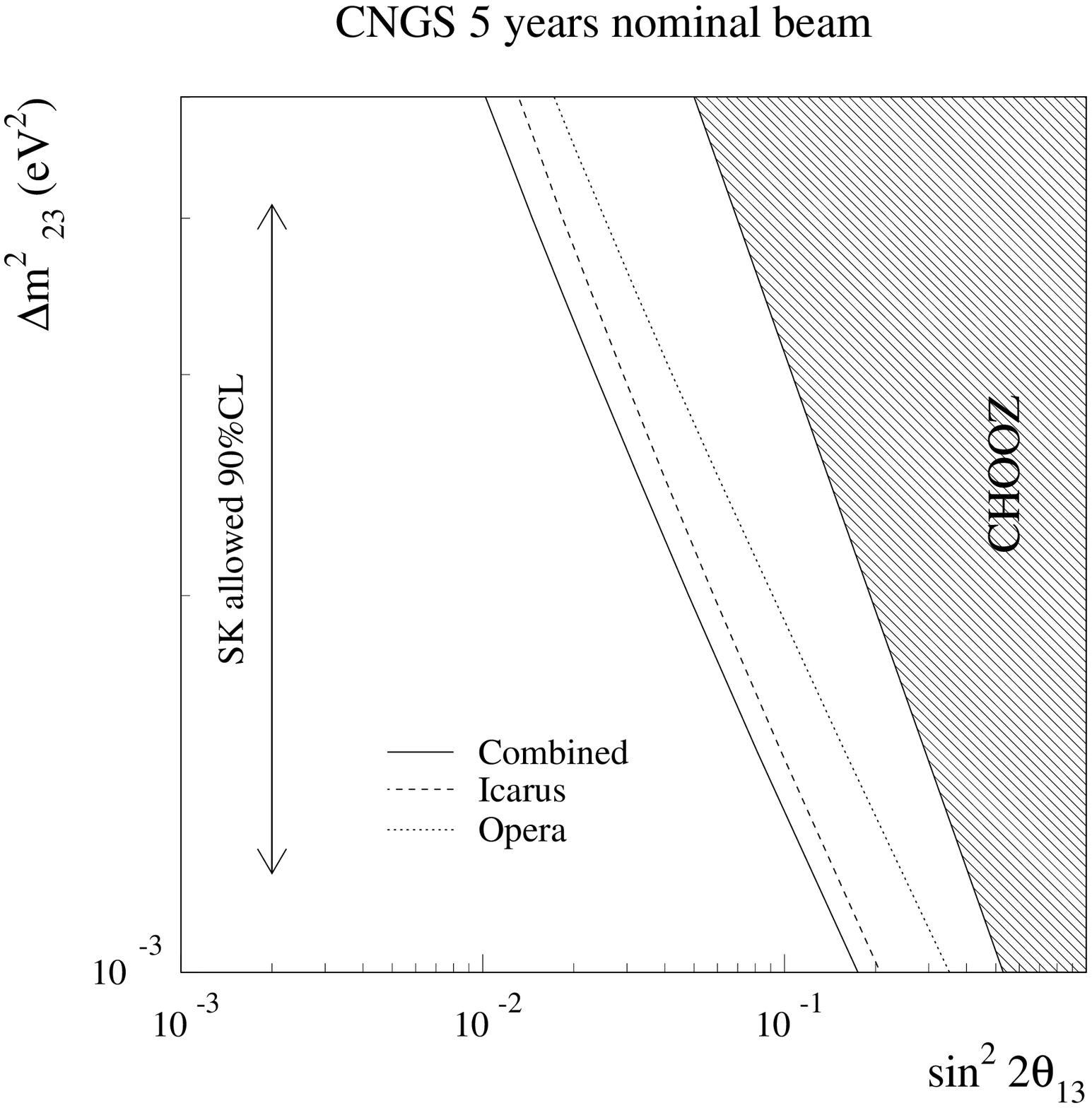,width=15cm}
\caption{Sensitivity to the parameter $\theta_{13}$ at 90\% C.L.in a
three family mixing scenario, in presence of
$\nu_\mu\leftrightarrow\nu_\tau$ with $\theta_{23}=45^\circ$.}
\label{fig:exclplot}
\end{figure}

By using the information contained in the Figs.~5.9~and~5.10 of
Ref.~\cite{icarus}, we have performed a binned $\chi^2$-fit of these
distributions and reproduced the results reported by the ICARUS Collaboration
in Ref.~\cite{icarus}. The sensitivity at 90\% C.L. to $\theta_{13}$,
corresponding to $\chi^2=\chi^2_{min}+4.6$, has been evaluated under the
assumption $\theta_{23}=45^\circ$. The exclusion plot is reported in
Fig.~\ref{fig:exclplot}.

\begin{figure}[htbp]
\centering
\epsfig{file=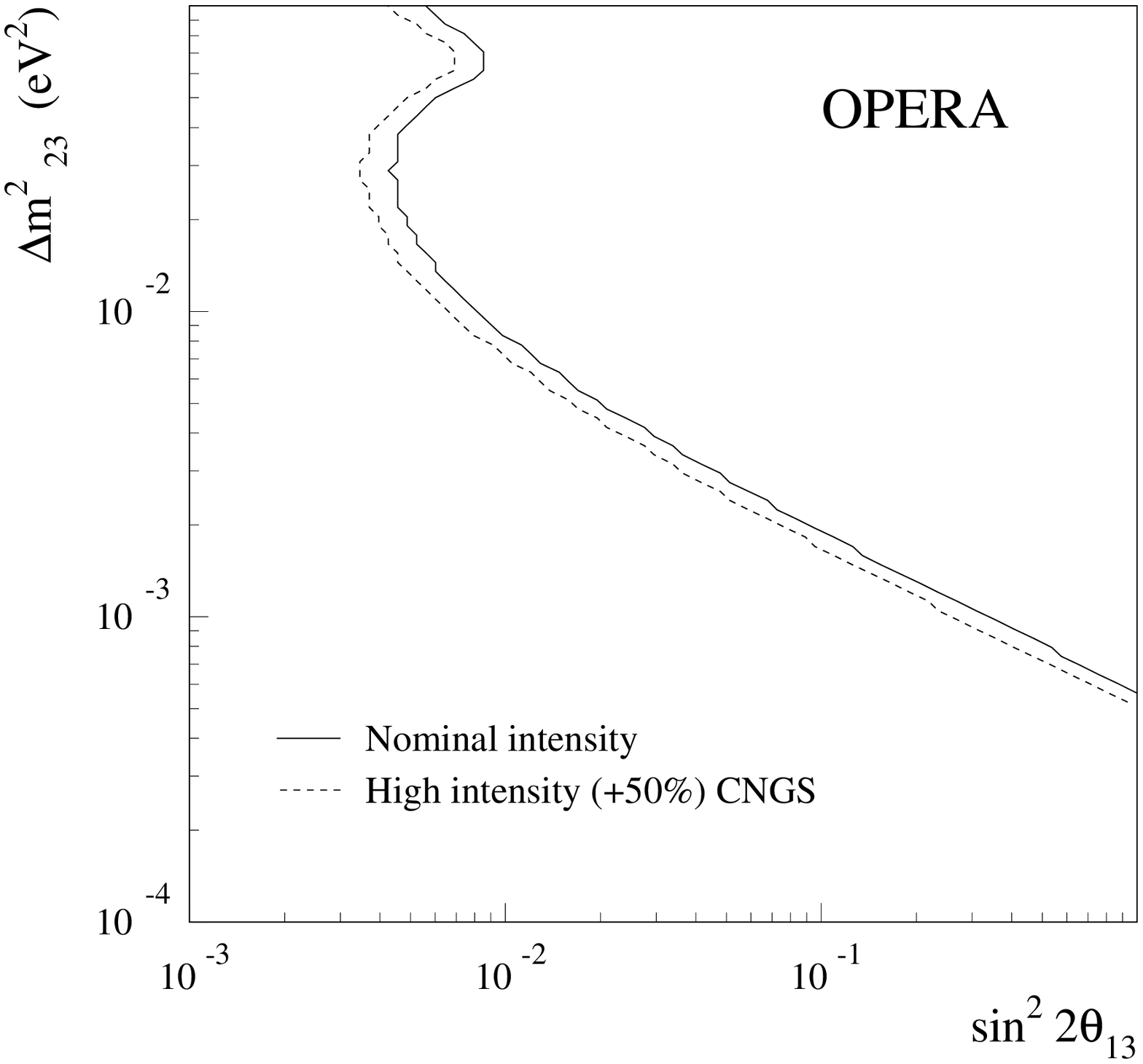,width=15cm}
\caption{OPERA sensitivity to the parameter $\theta_{13}$ at 90\% C.L. in a
three family mixing scenario, in presence of
$\nu_\mu\leftrightarrow\nu_\tau$ with $\theta_{23}=45^\circ$. The sensitivity
with the higher intensity CNGS beam is also given.}
\label{fig:exclplotopera}
\end{figure}

In the case of OPERA, the kinematical variables have been obtained starting
from true quantities and applying a smearing according to the OPERA
experimental resolutions, see Section~\ref{opera}. By fitting simultaneously
the $E_{vis},\, E_e,\, p_T^{miss}$ distributions , we obtained the exclusion
plot at 90\% C.L.  shown in Fig.~\ref{fig:exclplotopera} under the assumption
$\theta_{23}=45^\circ$.

The use of differential distributions improves the sensitivity to
$\sin^22\theta_{13}$ of both ICARUS and OPERA by about 20\%. The
combined sensitivity at 90\% C.L. of the ICARUS and OPERA experiments
is obtained by minimising the $\chi^2$ defined as
$\chi^2=\chi^2_{e(icarus)}+\chi^2_{e(opera)}$ and is also shown in
Fig.~\ref{fig:exclplot}.

\subsection{Results}

In the previous Sections we discussed the sensitivity achievable in
searching for $\theta_{13}$ with the present CNGS beam by combining
the ICARUS and OPERA experiments. The improvements on
$\sin^22\theta_{13}$ over the CHOOZ limit in the region indicated by
atmospheric neutrinos is about a factor 5 assuming 5 years data taking
with the nominal CNGS beam. The limits at 90\% C.L. on
$\sin^22\theta_{13}$ and $\theta_{13}$ for various experiments are
given in Table~\ref{tab:limits}. The limits achievable with an
improved CNGS beam (1.5 times more proton) are also shown.
 
It is worth noticing that the sensitivity of the combined experiments running
with the present high energy CNGS beam is comparable to the one achievable with
a dedicated low energy configuration running below the kinematical threshold
for $\tau$ production~\cite{Rubbia:2002rb} and is obtained ``for free'', as a
by-product of the search for $\nu_\tau$ appearance.

Clearly, we expect JHF phase 1 to improve significantly the
sensitivity to $\sin^22\theta_{13}$ with respect to the CNGS programme
in case of a negative result.

\begin{table}[hbtp]
  \small
\begin{center}
\caption{\small Limits at 90\% C.L. on $\sin^22\theta_{13}$ and $\theta_{13}$ for
various experiments.}     
\vspace{5mm}
\begin{tabular}{||c|c|c||}
\hline
\hline
Experiment & $\sin^22\theta_{13}$ & $\theta_{13}$ \\
\hline
CHOOZ & $<0.14$ & $<11^\circ$ \\
\hline 
MINOS & $<0.06$ & $<7.1^\circ$ \\
\hline 
ICARUS & $<0.04$ & $<5.8^\circ$\\
\hline 
OPERA & $<0.06$ & $<7.1^\circ$\\
\hline 
ICARUS and OPERA & & \\
combined (CNGS) & $<0.03$ & $<5.0^\circ$ \\
\hline
ICARUS and OPERA & & \\
combined (CNGSx1.5) & $<0.025$ & $<4.5^\circ$ \\
\hline
JHF~\cite{JHF} & $<0.006$ & $<2.5^\circ$\\
\hline 
\hline
\end{tabular}
\label{tab:limits}
\end{center}
\end{table}


\section{Conclusion}
\label{conclu}

The OPERA experiment was designed and optimised for the direct
observation of the $\tau$ decay topology and to search for $\nu_\tau$
appearance at the $\Delta m^2$ atmospheric scale. Nevertheless, given
the good electron identification, it turned out to be sensitive to the
sub-dominant $\nu_\mu \leftrightarrow \nu_e$ oscillations in the
region indicated by atmospheric results. Exploring the region of
$\sin^22\theta_{13}\simeq\mathcal{O}(10^{-2})$ is particularly
interesting in connection with the discovery of CP violation in the
leptonic sector~\cite{JHF}. Hence, an observation of
$\sin^22\theta_{13}$ in the region of $\mathcal{O}(10^{-2})$ would
place future CP violation searches like JHF phase 2 and the Neutrino
Factories on firmer ground. We showed that by using the present CNGS
beam and combining the ICARUS and OPERA experiments a sensitivity in
$\sin^22\theta_{13}$ down to 0.03 ($\Delta
m^2_{23}=2.5\times10^{-3}~\mbox{eV}^2$ and $\theta_{23}=45^\circ$) can
be achieved. It can be further improved if a higher intensity
($6.8\times 10^{19}$~pot/year) will be reached.


%
%

%
%

%
%

\end{document}